\documentclass[10pt]{article}
\usepackage[utf8]{inputenc}
\usepackage{epsfig}
\usepackage{amsmath,amsfonts,latexsym, amssymb}
\newtheorem{satz}{Theorem}[section]
\newtheorem{defi}[satz]{Definition}

\newtheorem{koro}[satz]{Corollary}

\newtheorem{assumption}[satz]{Assumption}

\newtheorem{conclusion}[satz]{Conclusion}
\newtheorem{ob}[satz]{Observation}

\newtheorem{postulate}[satz]{Postulate}
\newtheorem{conjecture}[satz]{Conjecture}

\newcommand{\mbf}{\mathbf}

\newcommand{\tit}{\textit}

\newcommand{\N}{\mathbb{N}}

\newcommand{\bewende}{$ \hfill \Box $}

\begin{document}
\thispagestyle{empty}
\begin{center}
\vspace*{1.0cm}

{\LARGE{\bf Wormhole Spaces: the Common Cause\\for the Black
    Hole Entropy-Area Law,\\the Holographic Principle and\\Quantum Entanglement }} 

\vskip 1.5cm

{\large {\bf Manfred Requardt }} 

\vskip 0.5 cm 

Institut f\"ur Theoretische Physik \\ 
Universit\"at G\"ottingen \\ 
Friedrich-Hund-Platz 1 \\ 
37077 G\"ottingen \quad Germany\\
(E-mail: requardt@theorie.physik.uni-goettingen.de)

\end{center}

\vspace{0.5 cm}

\begin{abstract}
  We present strong arguments that the deep structure of the quantum
  vacuum contains a web of microscopic wormholes or short-cuts. We
  develop the concept of wormhole spaces and show that this web of
  wormholes generate a peculiar array of long-range correlations in
  the patterns of vacuum fluctuations on the Planck scale. We conclude
  that this translocal structure represents the common cause for both
  the BH-entropy-area law, the more general holographic principle and
  the entanglement phenomena in quantum theory. In so far our approach
  exhibits a common structure which underlies both gravity and quantum
  theory on a microscopic scale. A central place in our analysis is
  occupied by a quantitative derivation of the distribution laws of
  microscopic wormholes in the quantum vacuum. This makes it possible
  to address a number of open questions and controversial topics in
  the field of quantum gravity.

\end{abstract} \newpage
\setcounter{page}{1}
%%%%%%%%%%%%%%%%%%%
\section{Introduction}
In the following we want to give a new explanation of the area law of
black hole (BH) entropy and the more general and stronger \tit{
  holographic principle}. Furthermore, we provide (in our view)
convincing arguments that an important structural ingredient of the
deep structure of our quantum vacuum is a network of microscopic
wormholes.  In contrast to e.g. \tit{string theory} and \tit{loop
  quantum gravity} (LQG), which both employ the quantum laws more or
less unaltered all the way down to the remote \tit{Planck scale}, we
regard this as an at least debatable assumption. We rather view the
holographic hypothesis as a means to understand how both quantum
theory and gravitation do emerge as derived and secondary theories
from a more fundamental theory living on a more microscopic scale. A
central role in this enterprise is played by an analysis of the
microscopic structure of the \tit{quantum vacuum} which leads to the
key concept of \tit{wormhole spaces}.

This important conceptual structure makes it possible to understand
the holographic aspects of quantum gravity, on the one hand, and the
(non-local) entanglement phenomena pervading ordinary quantum physics,
on the other hand, in a relatively natural way. Furthermore we think
that there exist links to the old ideas of e.g. Sakharov and
Zeldovich, dubbed \tit{induced gravity} (see for example \cite{Sa
  1},\cite{Sa 2},\cite{Z1},\cite{J1}).

Some words are in order regarding the relation of our investigation to
the analysis of BH entropy in, say, string theory. Three scenarios are
in our view in principle possible. Either, both approaches adress the
same phenomena in different languages, or they deal with them on
different scales of resolution of space-time. Be that as it may, we
think that our observation that the true ground state of our quantum
vacuum seems to be what we call a wormhole space (see section
\ref{worm}) is an aspect which is not apparent in the original string
theory approach and may be helpful to fix the proper ground state in
string theory.

In \cite{Bekenstein2} Bekenstein remarked that the deeper meaning of
black-hole entropy (BH-entropy) remains mysterious.
He asks, is it similar to that of ordinary entropy, i.e. the log of a
counting of internal BH-states, associated with a single BH-exterior?
(\cite{Bekenstein1},\cite{Bekenstein3} or \cite{Hawking1}). Or,
similarly, is it the log of the number of ways, in which the BH might
be formed. Or is it the log of the number of horizon quantum states?
(\cite{Hooft1},\cite{Susskind1}). Does it stand for information, lost
in the transcendence of the hallowed principle of unitary evolution?
(\cite{Hawking2},\cite{Giddings}). He then claims that the usefulness
of any proposed interpretation of BH-entropy depends on how well it
relates to the original ``statistical'' aspect of entropy as a measure
of disorder, missing information, multiplicity of microstates
compatible with a given macrostate, etc.

Quite a few workers in the field argue that the peculiar dependence of
BH-entropy on the area of the event horizon points to the fact that
the degrees of freedom (DoF), responsible for BH-entropy, are situated
near the event horizon. This seems to be further corroborated by the
corresponding behavior of the so-called entanglement entropy, i.e.
its (apparent) linear dependence on the area of the dividing surface
(cf., just to mention a few sources,
\cite{Hooft2},\cite{Sorkin1},\cite{Sorkin2} or the lively debate in
\cite{Jacobson1}, concerning entanglement entropy in a more general
setting, \cite{Sorkin3},\cite{Srednicki}). This linear dependence does
however not generally hold without further qualifications. It does in
particular \tit{not} hold for excited states (see \cite{Requ2})!

That is, while some particular sort of entanglement certainly plays an
important role in this context, the real question is in our view the
scale of resolution of space-time where this entanglement becomes
effective and the nature of the quantum vacuum on this level of
resolution.\\[0.2cm]
Remark: We want to emphasize the in our view crucial (but frequently
apparently not fully appreciated) point that the entropy content of a
BH is maximal.\\[0.2cm]
We think, the usual version of entanglement, we observe on the scales
of ordinary quantum theory, is only an \tit{epiphenomenon},
representing rather the coarse-grained effect of a hidden structure
which lives on a much more microscopic scale. I.e., we are sceptical
whether on such a microscopic scale the quantum vacuum can still be
treated in the way of an ordinary quantum field theory vacuum as
suggested in some of the papers cited above. We think, the
\tit{maximum-entropy property} of the BH-interior suggests another
interpretation. We will come back to this point in more detail in
section \ref{4} (cf. also the sceptical remarks in some of the review
papers by Wald, e.g. \cite{Wald1} (see in particular sect.6, Open
Issues), \cite{W2} (see in particular sect.4, Some unresolved Issues
and Puzzles],\cite{W3}).

As BH-entropy is widely regarded as an observational window into the
more hidden and primordial quantum underground of space-time, it
should be expected that it can be naturally explained within the
frameworks of the leading candidates of such a theory, i.e., to
mention the most prominent, string theory or LQG. For certain extreme
situations string theory manages to give an explanation of the
BH-entropy-area law. Whether the explanation is really natural is
perhaps debatable (it relies in fact on a number of assumptions and
correspondences as e.g. peculiar intersections of various classes of
p-branes). In a sense, it is rather a correspondence between
BH-behavior and the configurational entropy of certain string
states. To mention some representative papers,
\cite{S1},\cite{S2},\cite{S3},\cite{S4},\cite{S5},\cite{S6}.  In LQG,
on the other hand, it is assumed from the outset (at least as far as
we can see) that the corresponding DoF are sitting at the
BH-horizon. Therefore the observed area dependence of BH-entropy is
perhaps not so surprising (cf. e.g. \cite{L1},\cite{L2}).

In the enumeration of the most promising candidates for a theory of
quantum gravity one approach is usually left out which, we
nevertheless think, has a certain potential. One may, for example,
tentatively divide quantum gravity candidates into roughly three
groups, the relativisation of quantum theory (with e.g. LQG and causal
set theory as members), the quantisation of general relativity (string
theory being a prominent candidate) or third, theories which underlie
both general relativity \tit{and} quantum theory but are in fact more
fundamental and structurally different from both and contain these two
pillars of modern physics as derived and perhaps merely effective
sub-theories, living on coarser scales (cf. e.g. \cite{Isham}). In the
following we want to develop such a model theory in more detail.

As far as we can see, such a philosophy is also shared by `t Hooft who
emphasized this point in quite a few papers (see
e.g. \cite{H1},\cite{H2},\cite{H3},\cite{H4}). We quote from
\cite{H2}:
\begin{quote}{\small \ldots it may still be possible that the quantum
  mechanical nature of the phenomenological laws of nature at the
  atomic scale can be attributed to an underlying law that is
  deterministic at the Planck scale but with chaotic effects at all
  larger scales\ldots Since, according to our philosophy, quantum
  states are identified with equivalence classes\ldots}
\end{quote}
Furthermore:
\begin{quote}{\small\ldots It is the author's suspicion however, that these
  hidden variable theories failed because they were based far too much
  upon notions from everyday life and `ordinary physics' and in
  particular because general relativistic effects have not been taken
  into account properly.}
\end{quote}

While 't Hooft usually chooses his model theories from the cellular
automaton (CA) class, we are adopting a point of view which is on the
one hand more general and flexible but, on the other hand, technically
more difficult and complex. Instead of a relatively rigid underlying
geometric substratum in the case of CA (typically some fixed regular
lattice) on which the CA are evolving according to a given fixed
(typically local) CA-law, we are employing quite irregular, dynamic
geometric structures called by us \tit{cellular networks}, the main
point being that connections (\tit{edges} or \tit{links}) between the
respective \tit{nodes} or \tit{cells} can be created or annihilated
according to a dynamical law which, in addition, determines the
evolution of the local node- and edge-states.

To put it briefly, the `matter distribution' (i.e. the global pattern
of node-states) acts on the geometry of the network (the global
pattern of active edges) and vice versa. Thus, as in general
relativity, the network is supposed to find both its internal geometry
and its matter-energy distribution with the help of a generalized
dynamical law which intertwines the two aspects (cf. e.g. \cite{R1} or
\cite{R2} and further references given there). Technically, the
geometric substructure can be modelled by large, usually quite
irregular (\tit{random}) \tit{graphs}.

To make our point clear, this approach should not be confused with
e.g. the spin network approach in LQG or various forms of (dynamical)
triangulations. Our networks are usually extremely irregular and
wildly fluctuating on a microscopic scale, resembling rather Wheeler's
\tit{space-time foam}, and smooth geometric structures (as
e.g. dimensional notions) are hoped to emerge via some sort of a
\tit{geometric renormalisation process} (in fact a very particular
organized form of \tit{coarse-graining} steps). Some of the
interesting deeper mathematical aspects can for example be looked up
in \cite{R3}.

In our dynamical network approach to quantum space-time physics the
nodes are assumed to represent cells of some microscopic size
(presumably Planck size), the internal details of which cannot be
further resolved in principle or are ignored and averaged over for
convenience and will be represented instead by a simple ansatz for a
local (node) state. It can perhaps be compared with the many existing
spin-models which are designed to implement certain characteristic
features of complex solids. This is more or less the same philosophy
as in the CA-framework. The elementary connections between the nodes
(the edges in graph theory) are assumed to represent \tit{elementary
  interactions} or \tit{information channels} among the cells and also
carry simple \tit{edge-states}. We made a detailed numerical analysis
of the behavior of such networks in \cite{R4}.\\[0.2cm]
Remark: We would like to emphasize however, that our approach does not
really rely on this particular framework. It rather serves as a means
to illustrate the various steps in our analysis within a concrete
model theory.\vspace{0.2cm}

The paper is organized as follows. In the next section we analyse the
basic substratum, i.e. the microscopic patterns of vacuum
fluctuations, in particular the \tit{negative} energy fluctuations. In
section \ref{worm} we describe the three different roads which lead
(in our view: inevitably) to the concept of \tit{wormhole space}. The
preparatory sections 2 and 3 are then amalgamated in section 4 into a
detailed analysis of the microscopic distribution pattern of
\tit{short-cuts} or \tit{wormholes} and their consequences for the
number of \tit{effective} DoF in a volume of space. We introduce a new
type of dimension, the so-called \tit{holographic
  dimension}. Furthermore, we explain the microscopic basis of the
\tit{holographic principle} in general and the \tit{bulk-boundary
  correspondence} between the DoF in the interior of e.g. a BH and the
DoF on the boundary. Some apparent counter examples concerning the
area-scaling property (see e.g. \cite{Marolf}; Wheeler's 'bag of
gold'-spacetimes ) are very briefly addressed. In the last section we
briefly comment on a number of immediate applications of our
microscopic holographic approach and (open) problems which can be
settled with the help of our framework.

\section{The Structure of the Vacuum Fluctuations on a Microscopic Scale}
A characteristic feature of the dynamical network models we
investigated is their \tit{undulatory} character. As a consequence of
the \tit{feedback} structure of the coupling between node (cell)
states and \tit{wiring diagram} of edges (i.e. the pattern of
momentary elementary connections or interactions) the network never
settles in a static, frozen final state. The network may of course end
up in some \tit{attracting} subset of phase space but typical are wild
fluctuations on a small (microscopic) scale with possibly some
macroscopic patterns emerging on a coarser scale forming some kind of
\tit{superstructure} (see e.g. \cite{R4}).

It is in our view not sufficiently appreciated that, in contrast to
most of the other systems being studied in physics, the quantum vacuum
is in a state of eternal unrest on a microscopic scale, with, for all
we know, short-lived excitations constantly popping up and being
reabsorbed by the seething sea.

It therefore seems reasonable to regard our above network model
(investigated in e.g. \cite{R1} to \cite{R4}) as a (toy) model of the
quantum vacuum with the energy-momentum fluctuations on short scales
being associated with the fluctuations of the local node and edge
states.
\begin{postulate}In the following we adopt the working hypothesis of a
  parallelism of network behavior and microscopic behavior of the
  quantum vacuum.
\end{postulate}

We now come to a detailed analysis of the microscopic pattern of
vacuum fluctuations. In sect.4 of \cite{Requ1} we made a calculation
which shows that, given the huge number of roughly Planck-size grains
in a macroscopic piece of space and assuming that the individual
grains are allowed to fluctuate almost independently, more precisely,
some grain variable like e.g. the local energy, the total fluctuations
in a macroscopic or mesoscopic piece of space of typical physical
quantities are still so large (i.e. macroscopic) that they should be
observable. Note that with the number of nodes of roughly Planck-size,
$N_P$, in a macroscopic volume, $V$, being gigantic, its square root
is still very large (for the details of the argument see
\cite{Requ1}). More precisely, with $q_i$ some physical quantity
belonging to a microscopic grain (e.g.  energy, momentum, some charge
etc. and taking for convenience $\langle q_i\rangle$=0) and
$Q_V:=\sum_i q_i$ being the observable belonging to the volume $V$,
the fluctuation of the latter behaves under the above assumption as
\begin{equation}\langle Q_VQ_V\rangle^{1/2}\sim (V/l_p^3)^{1/2}\end{equation}
with $N_P\sim V/l_p^3$ the number of grains in $V$. This is a
consequence of the \tit{central limit theorem}. As such large
integrated fluctuations in a macroscopic region of the physical vacuum
are not observed (they are in fact microscopic on macroscopic scales),
we conclude:
\begin{conclusion}The individual grains or supposed elementary DoF do
  not fluctuate approximately independently.
\end{conclusion}
Remark: We note that this fact is also corroborated by other,
independent observations.\vspace{0.3cm}

We can refine the result further (cf. \cite{Requ1}) by assuming that
the fluctuations in the individual grains are in fact not independent
but correlated over a certain distance or, more precisely, are
\tit{short-range correlated}. In mathematical form this is expressed
as \tit{integrable correlations}.  This allows that ``positive'' and
``negative'' deviations from the mean value can compensate each other
more effectively. Letting e.g. $q(x)$ be the density of a certain
physical observable and $Q_V:=\int_V q(x)\,d^n\!x$ the integral over
$V$. In order that
\begin{equation}\langle Q_VQ_V\rangle^{1/2}\ll V^{1/2}
\end{equation}
we proved in \cite{Requ1} that it is necessary that
\begin{equation}\int_V \!d^n\!y\,\langle q(x)q(y)\rangle\approx 0 \label{fluc}   \end{equation}

We made a more detailed analysis in \cite{Requ1} under what physical
conditions property (\ref{fluc}) can be achieved, arriving at the
result:
\begin{conclusion}Nearly vanishing fluctuations in a macroscopic
  volume, $V$, together with short-range correlations imply that the
  fluctuations in the individual grains are anticorrelated in a
  fine-tuned and non-trivial way, i.e. positive and negative
  fluctuations strongly compensate each other which technically is
  expressed by property (\ref{fluc}).
\end{conclusion}
Remark: In \cite{R5} we extended such a vacuum fluctuation analysis
and applied it to measurement instruments, being designed to detect
(possibly) microscopic fluctuations of distances due to passing
gravitational waves.\vspace{0.2cm}

We hence infer that the fluctuation pattern of
e.g. energy-momentum has to be strongly \tit{anticorrelated}. But
under the above assumption it is possible that the underlying
compensation mechanism which balances e.g. the positive and negative
energy fluctuations is of short-range type, viz., individual
grain-energies may still fluctuate almost independently if their
spatial distance is sufficiently large. We show in the following that
the true significance of the so-called \tit{holographic principle} is
it, to enforce a very rigid and \tit{long-ranged} anticorrelated
fluctuation pattern in the quantum vacuum. As we are at the moment
only interested in matters of principle, we assume the simplest case
to prevail, called the \tit{space-like} holographic principle (holding
in contexts like e.g. quasi-static backgrounds or asymptotic
Minkowski-space; see e.g. the beautiful review \cite{Bousso}).
\begin{postulate}There exists a class of scenarios in which the
  maximal amount of information or entropy which can be stored in a
  spherical volume is proportional to the area of the bounding
  surface. The same holds then for the number of available DoF in
  $V$. This is the spacelike holographic principle.
\end{postulate}

In a series of papers Brustein et al. developed a point of view that
relates typical fluctuation results of quantum mechanical observables
in quantum field theory with the area-law-like behavior of
entanglement entropy and BH-entropy (cf. e.g. \cite{Bru}). We already
made a brief remark to this approach in \cite{Requ1}. We note
that we arrived at related results using different methods in another
context (see e.g. \cite{CMP50} and \cite{JMP43}). As a more detailed
comment would lead us too far astray, we plan to discuss this subject
matter elsewhere.

What we are going to show in the following is that the mechanism
leading to the strange area-behavior of the entropy of an enclosed
volume, $V$, is considerably subtler as usually envisaged. On the one
hand, we will show that the number of elementary DoF contained in $V$
is in principle proportional to the volume. On the other hand we infer
from observations on the macroscopic or mesoscopic scale that the
fluctuations of e.g. the energy are strongly anticorrelated. However,
as long as this compensation mechanism is short-ranged, we would still
have a number of nearly independently fluctuating clusters of
elementary DoF which again happens to be proportional to the volume as
the cluster size is roughly equal to the correlation length. So the
conclusion seems to be inescapable that the patterns of vacuum
fluctuations must actually be \tit{long-range correlated} on a
microscopic scale.

But we showed in \cite{Requ1} or \cite{R5} in quite some detail that
even systems, displaying long-range correlations, will usually have an
entropy which is proportional to the volume. A typical example is a
(quantum) crystal (\cite{Requ1},\cite{R5}). It is certainly correct
that below a phase transition point a system of particles in the
crystal phase has a smaller entropy than in the liquid or gas phase,
but still the entropy happens to be an extensive quantity. The reason
is in our view that the system develops, as a result of the long-range
correlations, new types of collective excitations (e.g. lattice
phonons) which serve as new collective DoF. Approximately they may be
treated as a gas of weakly interacting elementary modes with the usual
extensive entropic behavior.

That is, the holographic principle entails that the elementary DoF
have to be long-range anticorrelated (cf. also the remarks in
\cite{H4} or sect.7 of \cite{Susskind2}). But we see that this is only
a necessary but not a sufficient property for an entropy-area law to
hold. We hence arrive at the preliminary conclusion:
\begin{conclusion}From our preceding arguments and observations we
  conclude that the holographic principle implies that the fluctuation
  patterns in $V$ are long-range anticorrelated in a fine-tuned way on
  a microscopic scale and are essentially fixed by the state of the
  fluctuations on the bounding surface. The dynamical mechanism, which
  generates these long-range correlations must however, by necessity,
  have quite unusual properties (cf. subsection \ref{bulkboundary}).
\end{conclusion}
Before we derive the wormhole structure of the quantum vacuum on a
primordial scale in the next sections, we continue with the general
analysis of the pattern of vacuum fluctuations and derive some useful
properties of it.

A particular role is usually played by the energy and its
fluctuations. Furthermore, vacuum fluctuations are frequently
discussed together with the so-called \tit{zero-point energies}. While
they are not exactly the same, they are closely
related. Both occur also in connection with the \tit{cosmological
  constant problem} (to mention only a few sources see
e.g. \cite{Zinkernagel1},\cite{Zinkernagel2},\cite{Nernst},\cite{Enz},\cite{Boyer},\cite{Weinberg},\cite{Straumann}).

 In
the simplest examples like e.g. the quantized harmonic oscillator or the
electromagnetic field we have 
\begin{equation}H=P^2/2m+m\omega/2\cdot Q^2     \end{equation}
and with 
\begin{equation} \langle P\rangle_0=\langle Q\rangle_0 =0
\end{equation}
in the groundstate, $\psi_0$, we have
\begin{equation}\hbar\cdot\omega/2=\langle H\rangle_0=1/2m\cdot
  \langle (P-\langle P\rangle_0)^2\rangle_0+ m\omega/2\cdot\langle (Q-\langle Q\rangle_0)^2\rangle_0      \end{equation}
with 
\begin{equation}\langle (P-\langle P\rangle_0)^2\rangle_0\cdot \langle
  (Q-\langle Q\rangle_0)^2\rangle_0\geq \hbar^2/4   \end{equation}
which follows from $[P,Q]=-i\hbar$.
In the same way we have in (matter-free) QED:
\begin{equation}H=const\cdot (\mbf{E}^2+\mbf{B}^2)   \end{equation}
with
\begin{equation} \langle \mbf{E}\rangle_0=\langle \mbf{B}\rangle_0
  =0 \end{equation} so that again $\langle H\rangle_0$ is a sum over
pure vacuum fluctuations of the non-commuting quantities $\mbf{E}$ and
$\mbf{B}$.  One should however note that in the quantum field context products
of fields at the same space-time point have to be Wick-ordered (in
order to be well-defined). It is, on the other hand, frequently argued
that with gravity entering the stage, these eliminated zero-point
energy fluctuations have to be taken into account again. In our view,
this problem is not really settled.

We now come to an important point. It is our impression that in some
heuristic discussions (vacuum fluctuations as virtual
particle-antiparticle pairs) the consequences of the fact that
the vacuum state is an exact eigenstate of the Hamiltonian in a
Hilbert space representation of some quantum field theory are not
fully taken into account. I.e., we have 
\begin{equation}H\,\Omega=0 \end{equation}
(provided the ground state
energy is for convenience normalized to zero; note however that this
may be problematical in a theory containing gravity). Eigenstates,
however, have the peculiar property that the standard deviation is
necessarily zero,
\begin{equation}\Delta_{\Omega}\,H=\langle (H-\langle
  H\rangle_{\Omega})^2\rangle_{\Omega}^{1/2}=\langle H^2\rangle_{\Omega}^{1/2}=0      \end{equation}
According to the standard interpretation of quantum theory combined 
with spectral theory, $H^2\geq 0$, this implies that in each
individual observation process the total energy of the vacuum which
is, according to conventional wisdom, the (hypothetical) sum or
superposition of local (small scale) fluctuations, happens to be
exactly zero. In other words, the elementary fluctuations have to
exactly compensate each other in an apparently fine-tuned way. Put
differently
\begin{ob}If there are positive local energy fluctuations, there have
  to be at the same time by necessity negative energy fluctuations of
  exactly the same order. That is, at each moment, the global pattern
  of energy fluctuations in the quantum vacuum is an array of rigidly
  correlated positive and negative local excitations.
\end{ob}
Remark: Note the similarity of this independent observation to what we
have said above in connection with the holographic
hypothesis.\\[0.3cm]
It should be mentioned that Hawking in \cite{Haw1} invoked exactly
this picture of a particle pair excitation near the event horizon with
the virtual particle, having negative energy, falling into the BH
while the one with positive energy escapes to infinity.   

It would be useful to get more quantitative information on the
spectral properties of the local observables, in particular estimates
on negative fluctuations. One could try to make an explicit spectral
resolution of these quantities, e.g. of the energy, contained in a
finite volume, $V$, but this turns out to be difficult in general,
even if one has given an explicit model theory in some Hilbert
space. As we prefer a more general, model independent approach (not
necessarily based on Hilbert space mathematics), we proceed by using
(similar to Bell in his papers) a general probabilistic approach which
rather exploits the statistics of individual measurement
results. Unfortunately, we found that the standard estimates, known to
us in this context (e.g. the Markov-Chebyshev-inequality), always go
in the wrong direction (see e.g. \cite{Bauer} or
\cite{Feller}). Therefore we present in the following our own
estimate.

The strategy is the following. We take an observable, $E_V$, localized in $V$
with, for convenience, discrete spectral values, $E_i$, and
corresponding probabilities denoted by $p_i>0$. If we assume that the
expectation of $E_V$ is zero (which can always be achieved by a simple
shift) we have
\begin{equation}\sum\,p_i=1\quad , \quad \sum\,p_i\cdot
  E_i=0     \end{equation}
Furthermore, we assume its standard deviation in e.g. the vacuum,
$\Omega$, to be finite (which is automatically the case for bounded
operators, but we want to include also more general statistical variables)
\begin{equation}\sum\,p_i\cdot E_i^2=(\Delta_{\Omega}E)^2<\infty    \end{equation}

In a first step we make the simplifying assumption (taking e.g. a
bounded function of the energy)
\begin{assumption}
\begin{equation}|E_i|\leq\Lambda\quad\text{for all}\quad E_i    \end{equation}
\end{assumption}
We are interested in the amount of negative (e.g. energy) fluctuations
we will observe in measurements. A reasonable quantitative measure of
it is  
\begin{equation}\sum\,p_i^-\cdot |E_i^-|    \end{equation}
with $E_i^-,p_i^-$ the negative spectral values and their
corresponding probabilities. We then have (with $|E_i^-|/\Lambda\leq 1$)
\begin{equation}\sum\,p_i^-\cdot |E_i^-|/\Lambda\geq  \sum\,p_i^-\cdot |E_i^-|^2/\Lambda^2  \end{equation}
For the lhs we have
\begin{equation}\sum\,p_i^-\cdot |E_i^-|/\Lambda=\sum\,p_i^+\cdot
  |E_i^+|/\Lambda  \end{equation}
as the expectation of $E$ was assumed to be zero.

This yields
\begin{multline}\sum\,p_i^-\cdot |E_i^-|/\Lambda=1/2\cdot
  \left(\sum\,p_i^-\cdot |E_i^-|/\Lambda+\sum\,p_i^+\cdot
    |E_i^+|/\Lambda\right)\geq\\ 1/2\cdot
  \left(\sum\,p_i^-\cdot |E_i^-|^2/\Lambda^2+\sum\,p_i^+\cdot
    |E_i^+|^2/\Lambda^2\right)      \end{multline}
I.e.
\begin{equation}\sum\,p_i^-\cdot |E_i^-|\geq 1/2\cdot\sum\,p_i\cdot
  E_i^2/\Lambda=1/2\Lambda\cdot (\Delta_{\Omega}E)^2   \end{equation}
On the other hand (Cauchy-Schwartz)
\begin{equation}\left(\sum\,p_i^-\cdot |E_i^-|\right)^2=1/4\cdot
  \left(\sum\,p_i^-\cdot |E_i^-|+\sum\,p_i^+\cdot
    |E_i^+|\right)^2\leq 1/4\cdot\sum\,p_i\cdot |E_i|^2   \end{equation} 
We hence arrive at the result
\begin{conclusion}If the observable, $E$, is bounded, so that its
  spectral values fulfill $|E_i|\leq\Lambda$, we have the estimate
\begin{equation}1/2\Lambda^{-1}\,(\Delta_{\Omega}E)^2\leq
  \left(\sum\,p_i^-\cdot |E_i^-|\right)\leq 1/2\,
  (\Delta_{\Omega}E)      \end{equation} 
with $p_i$ the probabilities that the negative spectral values $E_i$
occur in an observation. That is, we manage to bound a quantity, which
is difficult to measure directly, by quantities, which are usually
more easily accessible.
\end{conclusion} 

We can generalize this result to situations where the $E_i$ are not
exactly bounded by some $\Lambda$ but are bounded in at least an
essential way. We assume that there exists some $\Lambda$ so that
\begin{equation}\sum_{|E_i|>\Lambda}\,p_i\cdot |E_i|^2<\varepsilon_{\Lambda}     \end{equation}
We then have
\begin{multline}\sum\,p_i^-\cdot |E_i^-|/\Lambda=1/2\,\left(\sum\,p_i^-\cdot |E_i^-|/\Lambda+\sum\,p_j^+\cdot |E_j^+|/\Lambda\right)\geq\\
1/2\,\left(\sum_{|E_i^-|\leq\Lambda}\,p_i^-\cdot |E_i^-|/\Lambda+\sum_{|E_j^+|\leq\Lambda}\,p_j^+\cdot |E_j^+|/\Lambda\right)\geq\\
1/2\,\left(\sum_{|E_i^-|\leq\Lambda}\,p_i^-\cdot |E_i^-|^2/\Lambda^2+\sum_{|E_j^+|\leq\Lambda}\,p_j^+\cdot |E_j^+|^2/\Lambda^2\right)\geq\\
1/2\,\left(\sum\,p_i\cdot
  |E_i|^2/\Lambda^2-\varepsilon_{\Lambda}/\Lambda^2\right)
\end{multline}
\begin{koro}Under the above assumption of an essentially bounded $E$ we have
\begin{equation}\sum\,p_i^-\cdot |E_i^-|\geq 1/2\Lambda\,\left((\Delta_{\Omega}E)^2-\varepsilon_{\Lambda}\right)     \end{equation}
\end{koro}

Another, rigorous, but not quantitative, argument can be derived from
axiomatic quantum field theory (see e.g. \cite{Wigh1}). It follows
from the so-called \tit{Reeh-Schlieder theorem} that there are no
\tit{local observables} or \tit{fields} which can annihilate the
vacuum (where by local we mean that the objects commute for space-like
separation). I.e., we have for any local $A$ (with $A=A^*$)
\begin{equation}A\,\Omega\neq 0\quad\Rightarrow\quad
  (A\Omega|A\Omega)=(\Omega|A^2\Omega)\neq 0 \end{equation}
 We take now as local observable the energy density integrated over a certain
spatial region, $V$,
\begin{equation}H_V:=\int_V\,h_{00}(\mbf{x},0)\,d^3\!x                \end{equation}
One usually normalizes $h_{00}(x)$ to
\begin{equation}(\Omega|h_{00}(x)\,\Omega)=0\quad\Rightarrow\quad (\Omega|\int_V\,h_{00}(\mbf{x},0)d^3\!x\,\Omega)=0       \end{equation}

The classical expression of the energy density, being derived in
Lagrangian field theory, is positive. The corresponding quantized
expression, after a necessary \tit{Wick-ordering} (see e.g. \cite{Bjo}
or \cite{Itz}) is however no longer positive definite as an operator
(density). This can be seen as follows. If the quantized energy
density were still positive, one can take the square root (via the
spectral theorem) of e.g. the positive operator $H_V$ and get:
\begin{equation}0= (\Omega|H_V\Omega)= (H_V^{1/2}\Omega|H_V^{1/2}\Omega)    \end{equation}
hence
\begin{equation}H_V^{1/2}\Omega=0    \end{equation}
As $H_V^{1/2}$ is also a local observable this is a contradiction due
to the Reeh-Schlieder theorem.
\begin{conclusion}$H_V$ is not a positive operator, hence its spectrum
  contains negative spectral values. It is then easy to construct
  Hilbert-space vectors, $\psi$, so that the measurement of $H_V$ in
  $\psi$ yields negative values for the local energy.
\end{conclusion}
Remark: We recently learned that this argument is originally
attributed to Epstein (unpublished;\cite{Reeh} or see
\cite{Summers1}), while the derivation which can be found in
\cite{Gla} is a completely different one.\vspace{0.2cm}

The important message (in our view) of all this is that, perhaps in
contrast to naive expectation, the quantum vacuum contains a lot of
negative energy excitations which globally exactly balance the
positive excitations. One may now speculate
about the possibility of making use of this observation.
%%%%%%%%%%%%%%%%%%%%%%%%
\section{\label{worm}Wormhole Spaces}
In this section we want to describe (very) briefly and sketchily the
three different lines of reasoning which lead us to the concept of
\tit{wormhole spaces}. The first line originated from our
investigation of the structure and dynamical behavior of the networks
we described above. In e.g. \cite{R1} we analyzed in some quantitative
detail the unfolding of the network structure and the various network
epochs under the inscribed microscopic dynamical laws and developed
the two-level concept of the network structure (or, rather, a
multi-scale structure), which, under the right conditions, is
relatively smooth on a sufficiently coarse-grained level (level 2)
with, among other things, a distant measure (metric) of the more
ordinary type and (hopefully) an integer-valued geometric dimension,
while on a more microscopic scale (level 1) the network structure is
expected to be very erratic with possibly a lot of links (elementary
interactions or information channels) connecting regions which may be
far apart with respect to the metric on level 2. The association of
these links with microscopic wormholes thus suggests itself (cf. in
particular observation 4.27 in \cite{R1}). Note furthermore that our
network dynamics implies that these translocal connections are
dynamically switched on or off. Compare this observation with the
point of view expounded in e.g. \cite{DeWitt}
\begin{quote}{\small \ldots But if a wormhole can fluctuate out of
    existence when its entrances are far apart \ldots then, by the
    principle of microscopic reversibility, the fluctuation \tit{into}
  existence of a wormhole having widely separated entrances ought to
  occur equally readily. This means that every region of space must,
  through the quantum principle, be potentially ``close'' to every
  other region, something that is certainly not obvious from the
  operator field equations which, like their classical counterparts,
  are strictly local.\ldots It is difficult to imagine any way in
  which widely separated regions of space can be ``potentially close''
to each other unless space-time itself is embedded in a convoluted way
in a higher-dimensional manifold. Additionally, a dynamical agency in
that higher-dimensional manifold must exist which can transmit a sense
of that closeness.}
\end{quote}

The quantitative network calculations in the mentioned papers have
mainly been performed within the framework of \tit{random
  graphs}. Important mathematical tools for the network analysis in
the transition from microscopic, strongly fluctuating and
geometrically irregular scales to coarse-grained and, by the same
token, smoother scales have been the concepts of \tit{cliques} of
nodes, the \tit{clique-graph} of a graph and an important network
parameter which we dubbed \tit{intrinsic scaling dimension} (we later
learned, \cite{R3}, that this concept plays also an important role in
\tit{geometric group theory} or \tit{Cayley-graphs} where it is called
the \tit{growth degree}).  To give a better feeling what is actually
implied, we give the definitions of clique, clique-graph and internal
scaling dimension (more about graph theory can e.g. be found in
\cite{Bollo}, \cite{R2}, notions and properties of graph dimension
were studied in e.g. \cite{ReqDim}).
\begin{defi}A simplex in a graph is a subset of vertices (nodes) with
  each pair of nodes in this subset being connected by an edge. In
  graph theory it is also called a complete subgraph. The maximal
  members in this class are called cliques.
\end{defi}
\begin{defi}The clique graph, $C(G)$, of a graph, $G$, is built in the
  following way. Its set of nodes is given by the cliques of $G$, an
  edge is drawn between too of its nodes if the respective cliques
  have a non-empty overlap with respect to their set of nodes.
\end{defi}

Graphs carry a natural neighborhood structure and notion of distance.
The neighborhood $U_n(x)$ of a node $x$ is the set of nodes $y$ which
can be reached, starting at $x$ in $\leq n$ consecutive steps,
i.e. there exists a path of $\leq n$ consecutive edges connecting the
nodes $x$ and $y$.
\begin{defi}\label{dist}The canonical network or graph metric is given by 
\begin{equation}d(x,y):=\min_{\gamma}\{l(\gamma)\,|\,\gamma\; \text{a path
connecting}\; x\; \text{and}\; y\}     \end{equation}
Here $l(\gamma)$ is the number of consecutive edges of the path. The
above definition fulfills all properties of a metric. Thus graphs and
networks are examples of \tit{metric spaces}.
\end{defi}
\begin{defi}[Internal Scaling Dimension]\label{Dim} 
  Let $x$ be an arbitrary node of $G$. Let $\#(U_n(x))$ denote
  the number of nodes in $U_n(x)$.We consider the sequence of
  real numbers $D_n(x):= \frac{\ln(\#(U_n(x))}{\ln(n)}$. We say
  $\underline{D}_S(x):= \liminf_{n \rightarrow \infty} D_n(x)$ is the
  {\em lower} and $\overline{D}_S(x):= \limsup_{n \rightarrow \infty}
  D_n(x)$ the {\em upper internal scaling dimension} of G starting
  from $x$. If $\underline{D}_S(x)= \overline{D}_S(x)=: D_S(x)$ we say
  $G$ has internal scaling dimension $D_S(x)$ starting from $x$.
  Finally, if $D_S(x)= D_S$ $\forall x$, we simply say $G$ has {\em
    internal scaling dimension $D_S$}.
\end{defi}
\begin{ob}We proved in \cite{ReqDim} (among other things) that this
  quantity does not depend on the choice of the base point for most
  classes of graphs.
\end{ob}
It turns out that this geometric notion is a very effective
characteristic of the large-scale structure of graphs and
networks. This topic was further studied in greater generality in
e.g. \cite{R3}.

In \cite{R2} we developed what we called the \tit{geometric
  renormalization group}, to extract important geometric coarse
grained, that is, large scale information from the microscopically
quite chaotically looking network and its dynamics. The idea is, at
least in principle, similar to the \tit{block spin transformation} in
statistical mechanics. That is, certain characteristic properties of
the system are distilled from the microscopically wildly fluctuating
statistical system by means of a series of algorithmic renormalization
steps (i.e. coarse-graining plus purification). The central aim is it
to arrive in the end at a system which resembles, on the surface, a
classical space-time, or, on the other hand, to describe the criteria
a network has to fulfill in order that it actually has such a
\tit{classical fixed point}.

In the course of this analysis we observed (cf. section VIII of
\cite{R2}) that the so-called \tit{critical network geometries},
i.e. the microscopic network geometries which are expected to play a
relevant role in the analysis, are necessarily in a very specific way
\tit{geometrically non-local}, put differently, they have to contain a
very peculiar structure of non-local links, or \tit{short-cuts}, that
is, in other words, the kind of \tit{wormhole structure}, we already
described above.

Relations to \tit{non-commutative geometry} were established and
studied in \cite{Connes}. We mention in particular section 7.2
``Microscopic Wormholes and Wheeler's Space-Time Foam'' and section 8
``Quantum Entanglement and Quantum Non-Locality''. The possible
relevance for quantum theory is in fact quite apparent (as has also
been emphasized in the papers by 't Hooft), as these microscopic
wormholes may be the origin of the ubiquitous entanglement phenomena
in quantum theory. The following figures describe pictorially the
nested structure of the cliques of nodes in consecutive
renormalization steps and overlapping cliques of nodes, defining the
local \tit{near-order} of \tit{physical points} together with shortcuts
which connect distant parts of the coarse-grained surface structure.
\begin{figure}[h]
\centerline{\epsfig{file=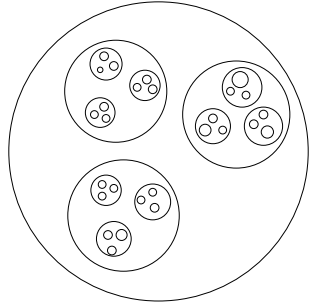,width=4cm,height=4cm,angle=0}}
\caption{Nested Structure; the (overlapping) cliques of a given level
  are represented as non-overlapping for reasons of pictorial clearness}
\end{figure}
\begin{figure}[h]
\centerline{\epsfig{file=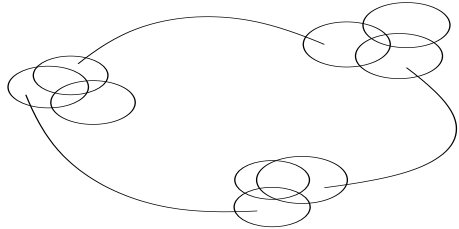,width=6cm,height=3cm,angle=0}}
\caption{Translocal links, connecting some local clusters of nodes (grains)}
\end{figure}

A second complex of (related) phenomena emerges in the field of
\tit{small world networks}. This is a particular class of networks of
apparently quite a universal character (described and reviewed in some
detail, for the first time, in \cite{Watts}) with applications in many
fields of modern science. They consist essentially of an ordinary
local network with its own local notion of distance superimposed by a
typically very sparse network of so-called \tit{short-cuts} living on
the same set of nodes and playing a structural role similar to the
microscopic wormholes described above. A typical example (with
dimension of the underlying lattice $k=1$) is given in the following
figure.
\begin{figure}[h]
\centerline{\epsfig{file=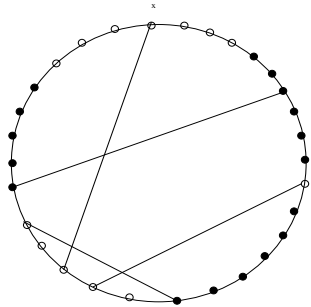,width=4cm,height=4cm,angle=0}}
\caption{The smallworld model for $k=1$. The number of nodes is
  $N=30$. In this particular realisation we have inserted four
  additional shortcuts. The unfilled nodes are the vertices which can
  be reached by for example $\leq 3$ steps starting from node $x$. The
  black nodes are the vertices not reached after three steps.}
\label{diagsmallworld}
\end{figure}
Some further (in fact very few) references, taken from quite diverse
fields are e.g. \cite{Strogatz},\cite{Grano},\cite{Lochmann}.
\begin{ob}Its, in our view, crucial characteristic is the existence of
  two metrics over the same network or graph. The first, $d_1(x,y)$,
  is defined (cf. definition \ref{dist}) by taking into account the
  full set of edges (i.e., including the short-cuts) and a second
  (local) metric, $d_2(x,y)$, taking into account only the edges of
  the underlying local network. It hence holds
\begin{equation}d_1(x,y)\leq d_2(x,y)    \end{equation}
\end{ob}
Remark: The metric $d_2(x,y)$ may then be associated (after some
renormalisation or coarse-graining steps) with an ordinary macroscopic
metric defined on a smooth space (without wormholes) like our
classical space-time. $d_1(x,y)$, on the other hand, should be
regarded as a microscopic distance concept which employs the existence
of wormholes.\vspace{0.2cm}

While, on the surface, the origin of this concept of small world
networks seems to be quite independent of the wormholes in general
relativity, it is the more surprising that on a conceptual meta level
various subtle ties do emerge. To mention only one (in our view)
important observation. In \cite{Schuster} it is for example shown,
that a sparse network of shortcuts superimposed upon an underlying
local network, has the propensity to stabilize the overall frequency
pattern (\tit{phase locking}) of so-called \tit{phase-oscillators}
which represent the nodes of the networks, the links representing the
couplings. The oscillators are assumed to oscillate with (to a certain
degree) independent frequencies. If we relate these local frequencies
with some local notion of time (or clocks), we may infer that
(microscopic) wormholes create or stabilize some global notion of
time!

We now come to the third strand, viz. the real wormholes of general
relativity or quantum gravity. We mainly concentrate on the wormholes
in true, i.e. Lorentzian space-time. Euclidean wormholes also (may)
play an important role and have been discussed extensively in the
context of the (nearly) vanishing value of the \tit{cosmological
  constant} (see e.g. \cite{HawW1},\cite{ColW},\cite{KlebaW},
\cite{PresW},\cite{UnrW},\cite{HawW2}). Of particular relevance in
the Lorentzian context are the so-called \tit{traversable}
wormholes. Their study started (as far as we know) with two seminal
papers by Thorne and coworkers (see \cite{Mo}). The geometric
construction of such solutions is in fact not so difficult if
performed by the so-called \tit{g-method}. That is, one constructs a
geometric wormhole, e.g. of the static type, and, in a second
step, computes the energy-momentum tensor being consistent with this
solution.

Giving a rough outline, this can be done in following way. Two open
balls are removed from two different pieces of e.g. approximately flat
3-space. Their boundaries are glued together with the junction being
smoothed. As a consequence of the smoothing process a tube emerges
interpolating between the two spheres (see e.g. \cite{Kras}). It is a
remarkable fact that in this process the \tit{weak energy condition}
(WEC) has to be violated, the latter implying that
\begin{equation}T_{00}\geq 0\quad ,\quad T_{00}+T_{ii}\geq
  0\quad\text{for}\quad i=1,2,3      \end{equation} 
that is, the matter-energy density is positive in any reference
system. Put differently, 
\begin{ob}In order to get a traversable wormhole, one has to violate
  the WEC. The WEC is always satisfied by classical matter. Therefore
  quantum effects are needed. The kind of negative energy needed is
  also called exotic matter.
\end{ob}

We showed in quite some detail in the preceding section that the
quantum vacuum abounds with negative energy fluctuations. Therefore
the speculation in section H of the first paper in \cite{Mo} does not
seem to be too far-fetched. In a next step one can study networks of
such traversable wormholes. In \cite{Ho1} it is speculated that such a
network, existing in the early universe, may solve the \tit{horizon
  problem}. The same situation was discussed from the point of view of
our network approach in section 4.1 (The Embryonic Epoch) of
\cite{R1}. All this comes already quite near the general picture we
envoked in the beginning of this section. Furthermore one can envisage
solutions combining black and white holes. This corresponds to some of
our networks where the orientation (direction) of the links connecting
two nodes can change under the dynamics. A review of Lorentzian
wormholes can be found in the book by Visser (\cite{Vi}). Some
other references are e.g. \cite{Red} and \cite{Few}. 

The above picture of a hypothetical network of wormholes sitting in
the deep structure of the quantum vacuum is beautifully complemented
by an approach (see e.g. \cite{Prep},\cite{Gara}) which investigates
within a (semi)classical approximation the energy of a quantum vacuum
state containing such an array of wormholes (or, rather, a gas of such
wormholes) and compare it with a vacuum state which in zeroth order is
flat Minkowski space. It comes out (apparently being a kind of Casimir
effect) that the quantum vacuum containing the wormhole gas has in
this semiclassical approximation a lower energy compared to the state,
being a perturbation of Minkowski space. One should note however that
this is a first order quantum effect! Anyhow, this observation seems
to corroborate the space-time foam picture of e.g. Wheeler and we
conclude this section with
\begin{conclusion}From our analysis in this and the preceding section
  emerges a model of the ground state of some preliminary version of
  quantum gravity which contains as an essential ingredient a network
  of microscopic wormholes. These wormholes can be created and
  annihilated and are in our picture the carriers of information
  between distant parts of classical space-time.
\end{conclusion}
\begin{defi}[Wormhole Space]We call such a physical structure a
  wormhole space and regard our cellular or small world networks,
  discussed above, as models, encoding and representing the typical
  characteristics of such systems. The typical characteristic is the
  existence of two types of distance, a microscopic one and an
  ordinary local one, being similar to ordinary macroscopic metrics on
  smooth spaces.
\end{defi}    
%%%%%%%%%%%%%%%%%%%%%%%%
\section{\label{4}Wormhole Spaces as the Common Cause of the Holographic Principle and the  Entropy-Area Law}
We learned in the preceding sections that two (presumably crucial)
properties govern the behavior of the quantum vacuum on a microscopic
scale. First, the vacuum fluctuations are strongly long-range
anticorrelated on a microscopic scale, i.e. there exists a fine-tuned
pattern of positive and negative (energy) fluctuations. Second, a
quantum mechanical stability analysis seems to show that the quantum
vacuum is pervaded by a network of microscopic wormholes. We argued
above that these two features are not independent phenomena but rather
are the two sides of the same medal. Furthermore, the presumed wormhole
structure has been supported by observations coming from
other fields of research like e.g. cellular or small-world networks.

In this (central) section we will now combine these observations and
show that they underlie (among other things) the \tit{holographic
  principle} and the \tit{entropy-area law} of BH-thermodynamics.  In
the following we will use (for convenience) the language of our
networks with the nodes of the network representing microscopic grains
of space (or space-time) of roughly Planck-size.  Leaving out other
details we treat our quantum vacuum as a wormhole space, i.e. as a
(small world) network consisting of an ordinary local network
structure being superimposed by a (presumably) sparse random network
with edges consisting of short-cuts, i.e. links, connecting regions of
space or space-time, which may be quite a distance apart with respect
to the metric, belonging to the underlying local network. These
short-cuts represent the wormholes of ordinary space-time.

The crucial characteristic, from which everything is expected to
follow, is the pattern and distribution of these short-cuts being
immersed in the underlying local network. That is, we randomly select
a node $x$ in the network $G$ ($G$ standing for graph) and study the
distribution of short-cuts connecting $x$ with nodes $y$ on spheres of
radius $R$ around $x$ (measured with respect to some macroscopic
metric or the natural metric of the underlying \tit{local}
network).
\begin{ob}We expect that the precise distribution law will depend on
  the concrete type of space-time we are dealing with. This holds in
  particular if the space-time is not static. That is, our microscopic
  approach to holography makes it possible to understand how
  holography may depend on the concretely given type of space-time
  (cf. e.g. the covariant entropy bound of Bousso, \cite{Bousso}).
\end{ob}
Remark: We emphasize that the network or the quantum vacuum it is
representing, is basically a statistical system with all local DoF
fluctuating. That means, most of our statements in the following are
about mean values or averages over finer statistical
details.
\subsection{The Distribution of Short-Cuts or Wormholes}
One can arrive at the law, describing the distribution of short-cuts
or wormholes around some arbitrary but fixed generic node (viz. some
fixed place in space-time) in roughly two ways. One can e.g.  motivate
the distribution law by appealing to certain fundamental principles
like e.g. \tit{scale-freeness} or absence of a particular and in some
sense unnatural length scale on a fundamental level. Alternatively,
one can show that a reasonable choice leads to far-reaching
consequences and corroborates the findings and observations made on a
more macroscopic level. To keep the discussion as briefly as possible
we adopt in this section the second point of view. In the following we
want to concentrate, for the sake of brevity, on a simple type of
quantum vacuum, that is, the vacuum belonging to ordinary Minkowski
space or a space-time which is asymptotically flat (e.g. a
Schwarzschild space-time). We postpone the analysis of more general
space-times as they occur in general relativity.

We make the following conjecture:
\begin{conjecture}On the average the number of short-cuts from a
  central node $x$ to nodes $y$, sitting on the sphere, $S_R(x)$ about
  $x$ is independent of $R$. Denoting this number by $N_{S_R}(x)$, we
  hence have
\begin{equation} N_{S_R}(x)=N_0           \end{equation}
\end{conjecture}
Remark: As this number is a statistical average, it need not be an
integer.\\[0.2cm]
The situation is depicted in the following picture. 
\begin{figure}[h]
\centerline{\epsfig{file=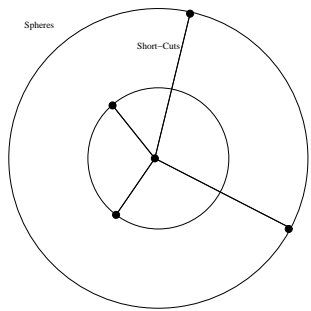,width=4cm,height=4cm,angle=0}}
\caption{Short-Cuts from a central node to nodes lying on two different
  spheres. In this picture we assumed $N_0=2$}
\end{figure}
\begin{ob}\label{bb} We will show in subsection \ref{bulkboundary} in a detailed
  quantitative analysis that this result approximately holds as
  well for nodes, not sitting exactly in the center of the spheres
  $S_R$ (see the following picture).
\end{ob}
\begin{defi}We denote the cluster of nodes in the ball $B_R$ being
  connected to an $x$ by short-cuts by $C_{B_R}(x)$.
\end{defi}
\begin{figure}[h]
\centerline{\epsfig{file=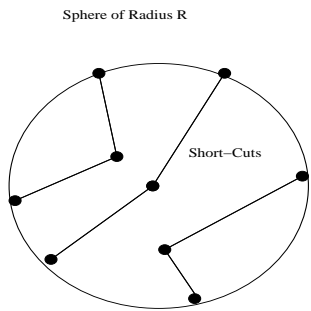,width=4cm,height=4cm,angle=0}}
\caption{Short-Cuts from nodes not sitting in the center to nodes
  lying on  a fixed sphere $S_R$.}
\end{figure}

We previously introduced the internal scaling dimension of a network
(see definition \ref{Dim}). It roughly describes how fast the network
is growing with respect to some base node. As this \tit{growth degree}
is to a large degree independent of the base node (see
e.g. \cite{ReqDim}) it is a global characteristic of a given network,
in fact of a whole class of similar networks (\cite{R3}). It is well
known that the generalization of the concept of dimension away from
smooth geometric structures is not unique. The above type of dimension
has the tendency to grow if additional short-cuts are inserted into a
given network geometry. We now introduce another dimensional concept
which catches other important network properties being more closely
related to the phenomena we want to analyze in this paper. It uses in
an essential way the two metrics, $d_1,d_2$, introduced above.
\begin{ob}From the above we infer that the number of nodes in the cluster
  $C_{B_R}(x)$ is approximately equal to $N_0\cdot R$. Furthermore,
  if the network of short-cuts is very sparse, the  clusters $C_{B_R}(x_i),C_{B_R}(x_j)$ with $x_i\neq x_j$ are
  essentially disjoint (the overlap is empty or very small). This is
  the phenomenon called {\em spreading} in the theory of random
  graphs.
\end{ob}
\begin{figure}[h]
\centerline{\epsfig{file=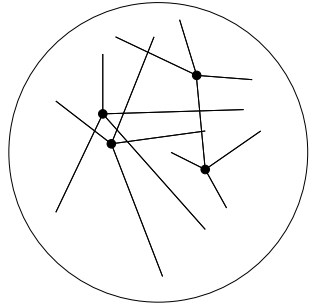,width=4cm,height=4cm,angle=0}}
\caption{Various clusters $C_{B_R}(x_i)$ with empty or marginal overlap}
\end{figure}
Hence, the following concept is reasonable.

We define a \tit{holographic dimension}, $D_H$, of a network in the
following way. We take some ball $B_R$ with macroscopic radius $R$
around some fixed but arbitrary node $x$ with respect to the local
metric $d_2$. We then form the $U_1^{(1)}(y)$-neighborhoods around the
nodes $y\in B_R$ with respect to the microscopic metric $d_1$
. We construct a \tit{minimal cover} of $B_R$ by such
$U_1^{(1)}(y_i)$, i.e. a minimal selection of such $y_i$ s.t.
\begin{equation}\bigcup_i\,U_1^{(1)}(y_i)\supset B_R    \end{equation}
The cardinality of such a minimal set we denote by $N_C(B_R)$. We take
the limit $R$ large or $R\to\infty$ (in an infinite network) and define 
\begin{defi}We call
\begin{equation}D_H:=\lim_{R\to\infty}\,\ln\,N_C(B_R)/\ln\,R     \end{equation}
the holographic dimension of the graph (network), provided the limit
exists. In the more general situation we can, as in definition
\ref{Dim}, define upper and lower dimensions etc.
\end{defi}
\begin{koro}As for the previously defined graph dimension, the limit
  is independent of the selected base point , $x$, if the network or
  graph is homogeneous on the average or in the large.
\end{koro}
\begin{ob}Due to the sparseness of the embedded subgraph of
  short-cuts, which yields the spreading property mentioned above, the
  number $N_C(B_R)$ scales for the wormhole spaces or small-world
  networks as  
\begin{equation}N_C(B_R)\sim R^{n-1}       \end{equation}
with $n$ the dimension of the local network or its coarse-grained
continuum limit space.
\end{ob}
Proof: The $U_1^{(1)}(y)$-neighborhoods consist of nodes lying in the
neighborhoods with respect to the local metric, $d_2$, $U_1^{(2)}(y)$,
plus the vertices connected by short-cuts with $y$. The cardinality of
$U_1^{(2)}(y)$ is independent of $R$ and typically (at least in our
models) a small number. For $R\to\infty$ $U_1^{(1)}(y)\cap B_R$ will
therefore consist mainly of nodes connected to $y$ by
short-cuts. Sparseness of the short-cut graph and spreading yield the
result.  \bewende
\begin{conclusion}For the type of wormhole spaces or small-world
  networks, defined above, we then have
\begin{equation}D_H=\lim_{R\to\infty}\,\ln(V(B_R)/R)/\ln\,R=n-1       \end{equation}
That is, in this case we have the important result
\begin{equation}D_H= dim\,S_R= n-1     \end{equation}
\end{conclusion}

We now come to the holographic principle and the BH-entropy area
law. As already mentioned, we discuss in this paper only the example
of 4-dim. asymptotically flat (Minkowski) space-time. In Planck units
a macroscopic ball, $B_R$, contains approximately 
\begin{equation}|V(B_R)|:= V(B_R)/l_p^3   \end{equation}
DoF or grains of Planck size. The typical cluster size is
\begin{equation}|C_{B_R}(x_i)|\approx N_0\cdot R/l_p     \end{equation}
Due to the mentioned spreading property the number of (effectively)
independent cluster in the above minimal cover is approximately
\begin{multline}N_C(B_R)\approx ((4/3)\pi\cdot R^3/N_0\cdot R)\cdot
  l_p^{-2}=(3N_0)^{-1}\cdot 4\pi R^2/l_p^2=\\(3N_0)^{-1}\cdot
  A(S_R)/l_p^2=:(3N_0)^{-1}\cdot |A(S_R)|            \end{multline}
with $A(S_R)$ denoting the area of $S_R$. 
\begin{ob}The number of effectively independent clusters,
  $C_{B_R}(x_i)$ in $B_R$ is  
\begin{equation}N_C(B_R)\approx (3N_0)^{-1}\cdot|A(S_R)|=
  (3N_0)^{-1}\cdot A(S_R)/l_p^2   \end{equation}
with the typical cluster size
\begin{equation}|C_{B_R}(x_i)|\approx N_0\cdot R/l_p       \end{equation}
\end{ob}

To show now that the number of effective DoF in a generic volume
(where by generic we mean a region in space with the diameter in all
directions being roughly of the same order) is proportional to the
surface area, $A(V)$, of its boundary, we employ a general
observation, made e.g. in statistical mechanics. An important tool for
the analysis of systems in statistical mechanics are correlation
functions. Correlations decay usually for large separation of the
respective DoF, but what is on the other hand certainly the case is,
that nearest neighbors are strongly correlated (near order versus far
order).
\begin{ob}We expect that the DoF in each of the $U_1^{(1)}(x)$ are
  strongly correlated. We hence take it for granted, that they act
  effectively as a single collective DoF.
\end{ob}
Remark: It may be possible, that this near order in the immediate
neighborhood of the grains can be finally destroyed by the insertion
of a huge amount of localized energy, but this does not seem possible
with present means. 
\begin{conclusion}[Area Law]Due to the existence of wormholes or
  short-cuts, distributed in space-time, the number of effective DoF
  (affiliated with the respective clusters $C_{B_R}(x_i)$) in e.g. a
  ball $B_R$ equals $N_C(B_R)$, that is
\begin{equation}\#(\text{DoF in}\;B_R)\approx (3N_0)^{-1}\cdot
  |A(S_R)|=(3N_0)^{-1}\cdot A(S_R)/l_p^2     \end{equation}
\end{conclusion}
This is the area-law behavior of entropy or number of DoF in a volume
of space found in e.g. BH-entropy. We note however, that this law, in
our formulation, is essentially a statement about the collective
behavior of the elementary DoF in (the interior of) a volume of
space. I.e., the respective DoF are \tit{not} really sitting on the
boundary of $V$. As to the details of the \tit{bulk-boundary
  correspondence} see the following subsection.

If we adopt the entropy-area law of BH-thermodynamics, which is,
expressed in Planck units,
\begin{equation} S=1/4\cdot |A| \end{equation} we have the possibility
to fix our parameter $N_0$, which gives the number of wormholes
connecting a central grain of space with the grains on a surrounding
sphere $S_R$ for any $R$. However, entropy is not exactly identical to
number of DoF. To relate the two, we have to make a simple model
assumption. One frequently makes the assumption of \tit{Boolean DoF},
i.e. the DoF on an elementary scale are \tit{two-valued}.
\begin{ob}With this assumption we have the relation
\begin{equation}S=N\cdot \ln\,2\quad\text{i.e.}\quad N=|A|/4\cdot\ln\,2            \end{equation}
with $S$ the entropy, $N$ the number of DoF.
\end{ob}
\begin{conclusion}With the help of this identification we get
\begin{equation}N_0=4/3\cdot\ln\,2   \end{equation}
which can in qualitative arguments be approximated by one!
\end{conclusion}

That is, in Planck units, there exists roughly one short-cut between a
central vertex and a surrounding sphere of radius $R$. This shows that
on an extremely microscopic scale, the network of short-cuts is indeed
very sparse. However the picture changes considerably if we go over to
more accessible length scales. If we use, for example an atomic
length-scale of e.g. $l_a:=10^{-10}m$, we have approximately 
\begin{equation}(10^{-10})^3/(10^{-35})^3=10^{75}    \end{equation}
grains of Planck-size in a volume element of diameter $l_a$. If we
then choose, instead of a sphere $S_R$, a spherical shell of radius
$R$ and thickness $l_a$ we have approximately
\begin{ob}The number of wormholes or short-cuts between a central
  volume element of size $l_a^3$ and a corresponding spherical shell
  of radius $R$ is approximately
\begin{equation}\#(\text{short-cuts})\approx 10^{75}\cdot 10^{25}=10^{100}        \end{equation}
which is quite a large number.
\end{ob}

If we choose for example $R=1m$, we see that roughly $10^{96}$ grains
in the shell are the endpoints of about $10^{100}$ short-cuts coming
from the central volume element of size $l_a^3$. If we replace $R$ by
the approximate diameter of the universe, i.e. $R_0\approx 10^{10}$ ly,
we get (with $1\,ly\approx 10^{17}m$):
\begin{equation}R_0\approx 10^{27}\,m     \end{equation}
and for the number of Planck-size grains in a spherical shell of this
radius:
\begin{equation}\#\,(\text{grains in shell of radius}\,R_0)\approx
  10^{149} \end{equation} with still $10^{100}$ short-cuts ending
there. That is, only one in $10^{49}$ grains is the endpoint of a
respective short-cut. But if we select a volume element of size
$l_a^3$ in this shell, we have still
\begin{ob}The number of wormholes (short-cuts) between two volume
  elements of size $l_a^3$  being a distance $R_0$ apart, is still
  the large number
  \begin{equation}\#\,(\text{short-cuts})\approx 10^{100}\cdot
    10^{-149}\cdot 10^{75}= 10^{-49}\cdot
    10^{75}=10^{26} \end{equation} that is, even over such a large
  distance there exist still a substantial number of wormholes
  connecting the two volume elements. But nevertheless, the network is
  sparse, viewed at Planck-scale resolution.
\end{ob}
\subsection{ \label{bulkboundary}The Bulk-Boundary Correspondence}
We now come to the last point of this section. From what we have
learned above, it is intuitively clear, that the DoF sitting on the
boundary $S_R$ of e.g. a ball $B_R$ should fix (or slave) the DoF in
the interior. But we note that in order that this can hold, we have to
verify our statement made in observation \ref{bb}. Furthermore, it is
of tantamount importance to understand in more quantitative detail the
influence of different shapes of the region under discussion and the
effect of different space-time geometries. The prerequisites for this
enterprise will be derived in the following.

As an example we employ, as we already did above, the simple geometry
of the spacelike holographic bound. For reasons of simplicity we place
the center of the ball in the origin, i.e. $\mbf{x_0}=\mbf{0}$. It is
of great help if we can transform the problem into a problem of
ordinary continuous analysis. To this end we introduce the probability
that a node in the interior of $B_R$ and an arbitrary node on the
boundary $S_R$ are connected by a short-cut. With $\mbf{y}\in S_R$ and
$\mbf{x}\in B_R$ there spatial euclidean distance in three dimensions
is
\begin{equation}|\mbf{y}-\mbf{x}|=\left(\sum_{i=1}^3\,(y_i-x_i)^2\right)^{1/2}   \end{equation}
\begin{ob}The edge probability is given by 
\begin{equation}p(|\mbf{y}-\mbf{x}|)=N_0/|A(S_{|\mbf{y}-\mbf{x}|})|=(N_0\cdot
  l_p^2/4\pi)\cdot |\mbf{y}-\mbf{x}|^{-2}
\end{equation}
\end{ob}
Here $|A(S_{|\mbf{y}-\mbf{x}|}|$ is the number of nodes (or
Planck-scale grains) on the sphere around
$\mbf{x}$ with radius $|\mbf{y}-\mbf{x}|$.This follows directly from
what we have learned in the previous sections. 

What we are actually doing in the following is the calculation of the
average number of short-cuts between an arbitrary node $\mbf{x}$ in
the interior of $B_R$ and the nodes on the boundary $S_R$. This will
be done within the framework of \tit{random graphs}. The above $p$ is
the so-called \tit{edge probability} (for the technical details see
\cite{Bollo} or \cite{R1},\cite{R2}). The sample space is the space of
graphs with \tit{node set} comprising the node in $\mbf{x}$ and all
the nodes sitting on the boundary $S_R$ and \tit{edge set} all
possible different sets of short-cuts connecting $x$ with the nodes on
$S_R$. The probability of each graph in the sample space is calculated
with the help of the above elementary edge probability $p$ and its
dual $q:=1-p$.

We choose $\mbf{x}$ arbitrary but fixed in $B_R(\mbf{0})$ and let
$\mbf{y}$ vary over the sphere $S_R(\mbf{0})$. The integral over
$S_R(\mbf{0})$ will then give the mean number of short-cuts between
$\mbf{x}$ and the grains on $S_R(\mbf{0})$. The guiding idea is that
the DoF in the interior are fixed by the DoF on the boundary if this
integral is essentially $\gtrsim 1$, as according to our philosophy,
developed previously, in that case every node in the interior has on
average at least one partner on the boundary as nearest neighbor with
respect to the microscopic metric $d_1$.

To make the integration easier we choose, without loss of generality, 
\begin{equation}\mbf{x}=\begin{pmatrix}0 \\ 0 \\ z \end{pmatrix}\quad
  , \quad z:=k\cdot R      \end{equation} 
with $0\leq k\leq 1$. A straightforward calculation (using polar
coordinates and appropriate variable transformations) yields for the
average number of short-cuts, $N_{S_R}(\mbf{x})$, 
\begin{multline}N_{S_R}(\mbf{x})= (N_0\,l_p^2/4\pi)\cdot
  l_p^{-2}\cdot\int_{S_R}\,|\mbf{y}-\mbf{x}|^{-2}\,d\!o=\\
\left(N_0/4\pi\cdot R^2\right)\cdot
2\pi\,R^{-2}\cdot\int_{-1}^{+1}\,d\!u\,((1+k^2)-2ku)^{-1}=\\N_0/2\cdot\int_{-1}^{+1}\,d\!u\,((1+k^2)-2ku)^{-1} \end{multline}
\begin{ob}Note that the integrand $((1+k^2)-2ku)^{-1}$ is always
  positive. Furthermore, our choice of a Coulomb-like law (in three
  dimensions) for the distribution of short-cuts in the previous
  subsection, i.e. $p\sim R^{-2}$, makes the above integral
  independent of $R$. 
\end{ob}

We can find a closed expression for the definite integral, i.e.
\begin{equation}I:=
  \int_{-1}^{+1}\,d\!u\,((1+k^2)-2ku)^{-1}=-1/2k\cdot\ln\,((1-k)^2/(1+k)^2)> 0   \end{equation}
Note that the position of the point $\mbf{x}$ relative to the center
and the boundary can be regulated by the value of the parameter $0\leq
k\leq 1$. We have tabulated the integral for $k$ from $0$ to $0.9$ in
the following table.\\[0.3cm]
\begin{tabular}{l|l|l|l|l|l|l|l|l|l|l}
$k$ & 0 & 0.1 & 0.2 & 0.3 & 0.4 & 0.5 & 0.6 & 0.7 & 0.8 & 0.9 \\ \hline
$I_k$ & 2 & 2 & 2 & 2.04 & 2.11 & 2.19 & 2.29 & 2.45 & 2.71 & 3.23
\end{tabular}\\[0.3cm]
We see that the number of short-cuts is almost constant through the
whole interior of $B_R$ apart from a thin shell near the boundary. But
this is not really surprising because there the main contribution
comes from the near side of the boundary and is no longer of a true
short-cut character. Taking into account the additional prefactor,
$N_0/2$, in front of the integral which is $\approx 1/2$ we have
\begin{conclusion}The number of short-cuts from an arbitrary node
  $\mbf{x}$ in $B_R$ to the boundary $S_R$ is approximately 
\begin{equation}p(\mbf{x})\gtrsim 1    \end{equation}
for most of the nodes. Furthermore for our Coulomb-like distribution
law it is independent of the radius of the sphere and is therefore
consistent with the expected holographic behavior for this geometry.
 \end{conclusion}

It is instructive to evaluate the above formula for $k>1$, i.e., the
influence via short-cuts of the sphere $S_R$ on a DoF in the exterior
of $S_R$. For $k$ large, the integral is dominated by the first term
in the integrand, viz. for $k$ large we have
\begin{equation}I\approx \int_{-1}^1d\!u\,(1+k^2)^{-1}\sim k^{-2}  \end{equation}
\begin{conclusion}For nodes, $x$, lying outside of $S_R$, the effect
  of the short-cut connections between $x$ and $S_R$ decays like a
  Coulomb-law. That is, the DoF in the exterior are no longer fixed by
  the DoF on $S_R$. What remains instead is a statistical influence in
  form of a correlation which decays with increasing distance. By the
  same token, there cannot be an entropy-area law for the exterior of
  the sphere relative to its internal boundary. Anyhow, this example
  does not really contradict the correctness of the spatial holographic
  principle as being presented in this paper. It would be interesting
  to relate our findings to the covariant holographic principle of
  e.g. Bousso, \cite{Bousso}
\end{conclusion}

This simple observation has an important consequence for arguments
being sometimes invoked against the general nature of the spatial
holographic principle (cf. e.g. \cite{Bousso}). While we do not intend
to discuss the holographic principle for more general space-times in
this paper, we mention one counter-example which one finds frequently
in the literature, i.e. a universe containing a closed spatial slice,
$S$ with a small inner subregion, $S_2$ (see the following picture).
\begin{figure}[h]
\centerline{\epsfig{file=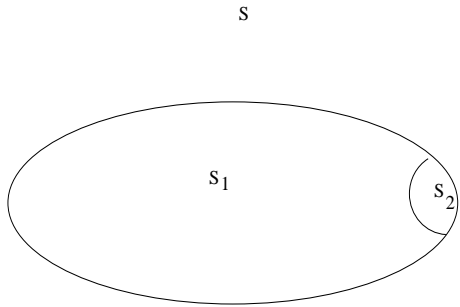,width=6cm,height=4cm,angle=0}}
\caption{A closed spatial slice containing a small subregion}
\end{figure}
\begin{ob}The area-law in the usual form applies for the subregion
  $S_2$ relative to its boundary. However, according to our
  (microscopic) version of spatial holography, the DoF on the inner
  boundary cannot slave the DoF in the large region $S_1$ if the inner
  boundary becomes too small. They only establish some kind of
  correlation in the exterior. The quantitative details are given by
  integrating our Coulomb-like influence law over the inner surface.
\end{ob}

Another, related, class of interesting (but perhaps pathological)
apparent counter examples (which we plan to address in greater detail
elsewhere) is discussed in e.g. \cite{Marolf}, i.e. spacetimes which
are called by Marolf 'bag-of-gold spacetimes'. An essential ingredient
is some FRW-spacetime hidden in the interior of a region which
resembles an ordinary BH. The innner FRW-universe has of course an
entropy which is proportional to its volume while from the outside the
whole configuration looks like a BH. This seeming contradiction can be
easily understood with the help of our microsopic holographic law as
the FRW-spacetime is actually only weakly coupled with the exterior of
the BH via wormholes. The technical arguments are the same as above.

\section{Commentary}
In the preceding sections we developed only the groundwork of our
approach. To keep the paper within reasonable size, we had to postpone
a more detailed discussion of the many consequences and immediate
applications. In this final section we at least undertake to briefly
comment on a number of important points. It is however obvious that a
more
detailed discussion of each point would require a paper of its own.\\[0.2cm]
i) The possible connections to the ubiquituous phenomenon of
entanglement in ordinary quantum theory are obvious. Interesting in
this respect is e.g. the well-known tension in quantum theory between
the locality and causality principle of special relativity and the
instantaneous state reduction, accompanying the measurement process
(cf. the respective sections in e.g. \cite{Aharonov}). We think,
similar to e.g. 't Hooft, that (the microscopic form of) holography
(we developed in this paper) is the common basis which may unite
quantum theory and
gravitation.\\[0.2cm]
ii) The consequences of the BH-entropy being maximal, which is quite
uncharacteristic for the ground state entanglement entropy in say
ordinary quantum theory, should be further
analysed.\\[0.2cm]
iii) The ADS-CFT-correspondence is regarded in string theory as the
paradigm for bulk-boundary correspondence (we mention only the review
\cite{Maldacena1} and the popular account \cite{Maldacena2}). In it
two, at first glance, fundamentally different theories are related to
each other, the one living in the bulk, the other living on the
\tit{boundary at infinity}. We must however say that the concrete
physical epistemology of this latter notion is not entirely clear to
us. The use of boundaries at infinity is wide spread in holography and
is mathematically well-defined, in particular for certain well-adapted
coordinate systems being in use in \tit{hyperbolic geometry}. But in
general it is rather an asymptotic property and not a concrete
place. Note that in our approach full information about the interior
of a (spatial) region is distributed essentially everywhere in the
exterior of the region via wormholes, but usually
not in the form of another field theory! \\[0.2cm]
iv) A virulent problem (the \tit{unitarity problem}) in
BH-thermodynamics is the question whether a pure state goes over into
a mixed state or not, that is, if the laws of ordinary quantum theory
are possibly violated in BH-thermodynamics (instead of the many
published papers we mention only the reviews by Wald, cited
above). This is a quite intricate epistomological problem somewhat
similar to the quantum measurement problem. We think, part of the
problem is that frequently pure states and mixtures are regarded as
complete opposites. But this is not really correct. It is here not the
place to go into more details. But in some respect it lies rather in
the eye of the beholder. That is, it is the problem of dealing with
the complete microscopic information of a state, or rather with some
coarse-grained form. Note that in our approach microscopic information
is widely scattered via short-cuts or wormholes over essentially the
whole space. I.e., it is not fully accessible to a local observer.  We
recommend the study of some older classics on the \tit{ergodic
  theorem} in quantum statistical
mechanics (\cite{Neumann},\cite{Pauli},\cite{Kampen}).\\[0.2cm]
v) Our analysis should be extended to more general space-times where
possibly different distribution laws may show up.

\end{document}